\documentclass{article}
\usepackage{amssymb}
\usepackage{amsmath}
\usepackage{graphicx}

\begin{document}

\title{On the origin of the minimal coupling rule, and on the possiblity of
observing a classical, \textquotedblleft
Aharonov-Bohm-like\textquotedblright\ angular momentum}
\date{White paper of 04-21-11 to Bong-Soo Kang}
\author{Raymond Chiao (rchiao@ucmerced.edu)}
\maketitle

Professor Roland Winston recently reminded me of Landau's argument in Landau
and Lifshitz's book \emph{The Classical Theory of Fields}  \cite{LXL}, which
starts from the principle of the relativistic invariance of the classical
action of a charged particle in the presence of classical electromagnetic
fields, whence one can derive the \textquotedblleft minimal coupling
rule\textquotedblright , viz.,%
\begin{equation}
\mathbf{p}\rightarrow \mathbf{p}-q\mathbf{A}  \label{minimal coupling rule}
\end{equation}%
and also the Lorentz force law%
\begin{equation}
\mathbf{F}=q\mathbf{E}+q\mathbf{v\times B}  \label{Lorentz force law}
\end{equation}%
This avoids the usual procedure of just postulating the minimal coupling
rule within standard quantum mechanics, or just adding the Lorentz force law
as an extra postulate to Maxwell's equations found in most textbooks on electrodynamics. Professor Winston emphasized to me how very powerful
relativistic invariance is: It puts such a powerful constraint on the form
of the action that it leads inexorably to both to the minimal coupling rule
and to the Lorentz force law. Since the minimal coupling rule is equivalent
to the principle of local gauge invariance, it is curious that one can
\textquotedblleft derive\textquotedblright , in some sense, the principle of
local gauge invarance from Lorentz invariance using Landau's argument. This
suggests that, in some hidden way, the symmetries of spacetime are the
source and origin of the local gauge principle, and therefore of all the
forces of nature.

Let us first start with the Feynman path-integral formulation of quantum
mechanics, which states that the probability amplitude of a particle to go
from point $a$ to point $b$ is the sum over all the probability amplitudes
for all possible paths joining points $a$ and $b$, i.e.,%
\begin{equation}
\left\langle b\right. \left\vert a\right\rangle =\sum_{n}{\exp \left( \frac{iS[\text{path 
}n]}{\hbar }\right)}  \label{Feynman sum over paths}
\end{equation}%
where the action, which is a functional, i.e., a function-like mathematical
object, which assigns a numerical value to an argument that is another
function $x_{n}(t)~$that specifies the path $n$%
\begin{equation}
S[\text{path }n]\equiv S[x_{n}(t)]
\end{equation}%
where%
\begin{equation}
S[\text{path }n]\text{=}\left. \int_{a}^{b}L(x,\dot{x})dt\right\vert
_{\text{path }n}  \label{S as integral of L dt}
\end{equation}%
where%
\begin{equation}
L(x,\dot{x})=T(\dot{x})-V(x)=\frac{1}{2}m\dot{x}^{2}-V(x)  \label{L = T- V}
\end{equation}%
is the Lagrangian evaluated along the given path $x_{n}(t)$, which need not
be the classical path, but can be any path joining the two fixed points $a$
and $b$, i.e., the known starting point $a$\ and the known ending point $b$\
of the particle. Point $a$ is to be interpreted as the point of emission of
the particle, and point $b$ is the point of detection of the particle.

The classical path is the one that extremizes the action, but Feynman
pointed out that in quantum mechanics, one must generalize the action to a 
\emph{sum} over paths, in order to include the contribution from all
possible paths, and not just the one (i.e., the classical path) that
extremizes the action. The superposition principle of quantum mechanics is
embodied in this sum. In fact, all possible paths must have the \emph{%
exactly equal weighting} specified by the phase-factor formula (\ref{Feynman
sum over paths}) in order to recover the standard Schr\"{o}dinger-equation
description for the evolution of the wavefunction. Extremizing the action is
equivalent, physically speaking, to the method of constructive interference
of Huygens' secondary wavelets; mathematically speaking, extremizing the
action is equivalent to the method of stationary phase.

It is clear from (\ref{Feynman sum over paths}) that the meaning of the
action $S[$path $n]$ (divided by the reduced Planck's constant) is that it
is the quantum mechanical phase $\phi $ which is accumulated by the particle
upon tranversing the path $n$ from point $a$ to point $b$, i.e.,%
\begin{equation}
\left. \phi \lbrack \text{path }n]\right\vert _{a}^{b}=\frac{\left. S[\text{%
path }n]\right\vert _{a}^{b}}{\hbar }
\end{equation}

Dirac, and later Feynman, introduced as a
postulate that a particle with a charge $q$ moving through a space filled by
a vector potential $\mathbf{A}$ will acquire an additional phase factor over
and beyond the usual phase factor arising from the action (\ref{S as
integral of L dt}) for a neutral particle with mass $m$ (\ref{S as integral
of L dt}). This extra phase factor is given by%
\begin{equation}
\exp (i\phi _{\text{charge}})=\exp \left( i\frac{q}{\hbar }%
\int_{a}^{b}\mathbf{A\cdot }d\mathbf{l}\right)
\label{Dirac phase factor}
\end{equation}%
where $d\mathbf{l}$ is a line element of the path taken by the charge. This
is equivalent to postulating that the Lagrangian has an extra piece $L_{%
\text{charge}}$\ in addition to (\ref{L = T- V}), viz.,%
\begin{equation}
L_{\text{charge}}=q\mathbf{A\cdot v}  \label{qv.A}
\end{equation}%
where $q$ is the charge of the particle, $\mathbf{v}$ is the three-velocity
of the charged particle moving through a region of space with a vector
potential $\mathbf{A}$. From the extra piece of the Lagrangian (\ref{qv.A}),
it follows that, at the quantum level of description, there will exist an
extra piece of \textquotedblleft charged\textquotedblright\ momentum gotten
by taking a partial derivative of the Lagrangian with respect to the
velocity $\mathbf{\dot{x}=v}$%
\begin{equation}
\mathbf{p}_{\text{charge}}=\frac{\partial L_{\text{charge}}}{\partial 
\mathbf{\dot{x}}}=\frac{\partial L_{\text{charge}}}{\partial \mathbf{v}}=q%
\mathbf{A}
\end{equation}%
which we shall call this the \textquotedblleft $q\mathbf{A}$%
-type\textquotedblright\ momentum, or \textquotedblleft electromagnetic
momentum\textquotedblright , which a particle with a charge $q$ would
possess by virtue of its charged coupling to electromagnetic fields, which
is in addition to the standard \textquotedblleft $m\mathbf{v}$%
\textquotedblright\ momentum, or \textquotedblleft
kinetic\textquotedblright\ momentum, that a neutral particle with mass $m$
moving with a velocity $\mathbf{v}$\ would possess. Note that even if the
charged particle $q$ is at rest, so that its \textquotedblleft $m\mathbf{v}$%
\textquotedblright\ momentum is zero, it can still possess a non-zero
\textquotedblleft $q\mathbf{A}$-type\textquotedblright\ momentum in the
presence of a non-vanishing $\mathbf{A}$\ field, such as that arising from
the flux inside an infinitely long solenoid in the Aharonov-Bohm effect.

But where does this extra piece of the Lagrangian $q\mathbf{A\cdot v}$ come
from? Why does the vector potential appear here instead of the more
\textquotedblleft physically real\textquotedblright\ electric or magnetic
fields? What happened to gauge invariance here? Does the extra piece of the
Lagrangian $q\mathbf{A\cdot v}$ exist at the classical level of description
of the motion of a classical charged particle in the presence of classical
electromagnetic fields, as well as at the quantum level?

In order to answer these questions, let's return to an entirely \emph{%
classical} description of the motion of a charged particle in a \emph{%
classical} electromagnetic field, and let's follow Landau's method of
constraining the form of the classical action for the charged particle's
motion based on the principle of Lorentz invariance, following Professor
Winston's suggestion. Let's first generalize (\ref{S as integral of L dt})
to its relativistic form%
\begin{equation}
S[\text{path }\gamma ]\text{=}\left. \int_{a}^{b}L(x^{\mu },\frac{%
dx^{\mu }}{d\tau })d\tau \right\vert _{\text{path }\gamma }
\end{equation}%
where $x^{\mu }$ is the four-vector position of a classical particle and%
\begin{equation}
v^{\mu }=\frac{dx^{\mu }}{d\tau }
\end{equation}%
is the four-velocity of the particle, where the spacetime path $\gamma $ is
that of the classical trajectory of a charged particle from the starting
spacetime point $a$ to the ending spacetime point $b$ of the trajectory in
spacetime, and where $d\tau $ is the infinitesimal increment of proper time
of the particle along its trajectory. Landau demanded that the form of the
action $S$ must be relativistically invariant under all possible Lorentz
transformations, i.e., it must have the form of a four-scalar, or invariant,
in spacetime.

Now let the particle have both a mass $m$ and a charge $q$. Then the total
action (where we omit the limits of integration and suppress the path
specification as being understood) will be the sum of two parts%
\begin{equation}
S=S_{\text{mass}}+S_{\text{charge}}
\end{equation}%
where the first term must have the relativistically invariant form%
\begin{equation}
S_{\text{mass}}=mc^{2}\int d\tau =mc\int ds  \label{S_mass}
\end{equation}%
where $m$ is the rest mass of the particle, where $c$ is the speed of light,
and where the invariant interval $ds$ in general relativity is defined
through a quadratic form via the metric tensor $g_{\mu \nu }$%
\begin{equation}
ds^{2}=g_{\mu \nu }dx^{\mu }dx^{\nu }  \label{metric tensor}
\end{equation}%
(We shall use Einstein's summation convention for repeated spacetime
indices, i.e., for Greek-letter spacetime indices running from 0 to 3, with
a metric signature of ($-$1, +1, +1, +1)). In the case of special relativity%
\begin{equation}
ds^{2}=\eta _{\mu \nu }dx^{\mu }dx^{\nu }
\end{equation}%
where the Minkowski tensor is the diagonal tensor defined as follows:%
\begin{equation}
\eta _{\mu \nu }=\text{diag}\left( -1,+1,+1,+1\right)
\end{equation}%
so that using Cartesian coordinates%
\begin{equation*}
ds^{2}=-c^{2}dt^{2}+dx^{2}+dy^{2}+dz^{2}
\end{equation*}

Note that in special relativity, $S_{\text{mass}}$ is manifestly a Lorentz
scalar, and is therefore relativistically invariant. Furthermore, it has the
units of the product of energy and time, i.e., the units of Planck's
constant, so that $S_{\text{mass}}/\hbar $ is a dimensionless quantity,
which is consistent with the quantum phase $\phi =S_{\text{mass}}/\hbar $
being dimensionless. Note that even if the particle were to be neutral,
i.e., with $q=0$, but $m\neq 0$, it still must have a contribution $S_{\text{mass}%
}$ to its action. This term, when extremized using the standard variational
principle $\delta S=0$, yields, in the non-relativistic limit, Newton's
second law of motion via the standard Euler-Lagrange equations of motion.
Thus one recovers the standard nonrelativistic form of classical mechanics.

Next, let's consider all the possible forms for $S_{\text{charge}}$ that are
allowed by relativistic invariance. The action $S_{\text{charge}}$ at the
classical level represents a measure of the coupling of the motion of a charged
particle to all classical electromagnetic fields, and at the quantum level
it becomes the phase change of the particle arising from the interaction of
a moving charge with these fields. The motion of the charged particle is
described by the four-vector%
\begin{equation}
v^{\mu }=\frac{dx^{\mu }}{d\tau }=\text{particle's four-velocity}
\end{equation}%
In a proper-time increment $d\tau $, the particle is displaced by the
four-vector%
\begin{equation}
dx^{\mu }=v^{\mu }d\tau =\text{particle's four-displacement}
\end{equation}%
We seek an action $S_{\text{charge}}$ which satisfies the \emph{linearity}
requirement%
\begin{equation}
dS_{\text{charge}}\propto dx^{\mu }  \label{S linear in dx}
\end{equation}%
for infinitesimal increments of $dx^{\mu }$ in the presence of
electromagnetic fields. This requirement follows from the fact that all
\textquotedblleft physically reasonable\textquotedblright\ fields must
become \emph{uniform} fields, when they are viewed on the tiny,
infinitesimal length scales given by $dx^{\mu }$. Hence the action of
displacing a charge in the presence of such uniform fields by an
infinitesimal amount $dx^{\mu }$ must be \emph{linear} in $dx^{\mu }$, with
the exception of the gravitational field, for which a uniform gravitational
field, such as the $\mathbf{g}$ field of the Earth, acting on a particle
within an infinitesimally small region of spacetime, can always be
transformed away by the equivalence principle. Therefore, instead of a \emph{%
linear} dependence on $dx^{\mu }$ given by $dS_{\text{charge}}\propto
dx^{\mu }$, one must use for the action of gravity on a particle the \emph{%
square-root of the quadratic form} given by%
\begin{equation}
dS_{\text{mass}}\propto \sqrt{g_{\mu \nu }dx^{\mu }dx^{\nu }}
\label{square root}
\end{equation}%
Note the contrast between the \emph{non-analytic} nature of the square-root
function in (\ref{square root}), versus the \emph{analytic} nature of the
linear function in (\ref{S linear in dx}). (I thank Steve Minter for
pointing out this imporant difference to me).

Furthermore, we shall require that the action%
\begin{equation}
dS_{\text{charge}}\propto q
\end{equation}%
be directly proportional to the size and the sign of the charge of the
particle which is interacting with an electromagnetic field. We shall assume
here that the charge $q$ is a Lorentz invariant quantity. It follows that%
\begin{equation}
dS_{\text{charge}}\propto qdx^{\mu }
\end{equation}%
However, we note that $dS_{\text{charge}}$ transforms as a four-scalar,
whereas $qdx^{\mu }$ transforms as a four-vector, under Lorentz
transformations. The only way that this can happen is if we contract $%
dx^{\mu }$ with some other four-vector. The only four-vector that one can
reasonably associate with electromagnetic fields is the vector potential $%
A_{\mu }$. This suggests that we try%
\begin{equation}
dS_{\text{charge}}=qA_{\mu }dx^{\mu }  \label{Trial solution for S_charge}
\end{equation}%
as an Ansatz for the action of electromagnetic fields acting on a charge $q$%
. This trial solution for $dS_{\text{charge}}$\ has the right dimensions (we
use SI units here).

There remains a sign ambiguity in this Ansatz, however, which is resolved by
taking the nonrelativistic limit, in which case one gets%
\begin{equation}
dS_{\text{charge}}=+q\mathbf{A\cdot v}dt=L_{\text{charge}}dt
\label{NR limit of charged action}
\end{equation}%
which has the correct sign given the usual sign conventions for charges,
currents, and magnetic fields (i.e., \emph{positive} charges as the carriers
of \emph{positive} electrical currents are the sources of magnetic fields
via the usual \emph{right-hand} rule), and given Landau and Lifshitz's
metric signature of ($-$1, +1, +1, +1). Thus one recovers the
non-relativistic charge Lagrangian of the form (\ref{qv.A}) with the correct
sign, and the minimal coupling rule (\ref{minimal coupling rule}) also with
the correct sign. One also recovers the Lorentz force law (\ref{Lorentz
force law}) with the correct sign.

It follows that, at the classical level of description, the non-relativistic
limit (\ref{NR limit of charged action}) of Landau's relativistic classical
action (\ref{Trial solution for S_charge}) leads to the conclusion that the
charge $q$ must have an extra piece of \emph{classical} \textquotedblleft
electromagnetic\textquotedblright\ momentum%
\begin{equation}
\mathbf{p}_{\text{charge}}=\frac{\partial L_{\text{charge}}}{\partial 
\mathbf{\dot{x}}}=\frac{\partial L_{\text{charge}}}{\partial \mathbf{v}}=q%
\mathbf{A}
\end{equation}%
or a \textquotedblleft $q\mathbf{A}$-type\textquotedblright\ momentum, in
addition to the usual, classical \textquotedblleft $m\mathbf{v}$%
\textquotedblright\ momentum of a neutral particle with mass $m$, whenever
and wherever a classical charge $q$ exists in the presence of a
non-vanishing classical vector potential $\mathbf{A}$. A classical charged
particle possesses this extra piece of momentum by virtue of its charged
coupling to electromagnetic fields. It should be emphasized that this extra
piece of momentum $\mathbf{p}_{\text{charge}}=q\mathbf{A}$ exists in the 
\emph{classical} problem of the motion of the charge $q$, and not only in
the \emph{quantum} problem of the motion of $q$.

Luis Martinez then asked: Why can't one perform a gauge transformation on
the vector potential $\mathbf{A}$ so that it is zero at all points in space,
i.e.,%
\begin{equation}
\mathbf{p}_{\text{charge}}=q\mathbf{A=0}
\end{equation}%
everywhere? For it is always possible to choose a gauge
transformation such that at every point in space%
\begin{equation}
\mathbf{A\rightarrow A+\nabla }\chi =\mathbf{0}
\end{equation}%
by an appropriate choice of the arbitrary scalar function $\chi $ at each
point in space. Then there would never exist any such thing as a
\textquotedblleft $q\mathbf{A}$ -type\textquotedblright\ or an
\textquotedblleft electromagnetic\textquotedblright\ momentum of the charge $%
q$, since it can always be transformed away.

However, as in the quantum Aharonov-Bohm effect, whenever there is a $closed$
curve $C$ that encloses a solenoid with nonvanishing magnetic flux within
it, then, by Stokes's theorem,%
\begin{equation}
\oint_{C} \mathbf{A\cdot }d\mathbf{l}= \iint_{S\left( C\right)
}\left( \mathbf{\nabla \times A}\right) \mathbf{\cdot }d\mathbf{S=}%
\iint_{S\left( C\right) }\mathbf{B\cdot }d\mathbf{S}=\Phi _{\text{enc}}\neq
0
\end{equation}%
which cannot be transformed away by any gauge transformation, since $\Phi _{%
\text{enc}}$ is a gauge-invariant quantity. Note that this argument holds at
both the classical and quantum levels of description of the motion of a
charge $q$.

One may then ask: Why not try coupling the charge directly to the more
\textquotedblleft physically real\textquotedblright\ electric and magnetic
fields via the electromagnetic field tensor \cite{Number of components}%
\begin{equation}
F_{\mu \nu }=\frac{\partial A_{\mu }}{\partial x^{\nu }}-\frac{\partial
A_{\nu }}{\partial x^{\mu }}
\end{equation}%
which has the advantage of being manifestly gauge invariant? For example,
one might try%
\begin{equation}
dS_{\text{charge}}^{\prime }=qF_{\mu \nu }dx^{\mu }dx^{\nu }
\end{equation}%
The problem with this alternative \textquotedblleft trial\textquotedblright\
four-scalar solution is two-fold: (1) $dS_{\text{charge}}^{\prime }\propto
dx^{\mu }dx^{\nu }$ is \emph{quadratic} in $dx^{\mu }dx^{\nu }$, and
therefore violates the \emph{linearity} requirement (\ref{S linear in dx});
(2) $F_{\mu \nu }$ is antisymmetric in the exchange of the two indices $\mu $
and $\nu $, but $dx^{\mu }dx^{\nu }$ is symmetric. Hence by symmetry%
\begin{equation}
dS_{\text{charge}}^{\prime }=qF_{\mu \nu }dx^{\mu }dx^{\nu }=0
\end{equation}%
Hence it is natural to call Landau's trial solution (\ref{Trial solution for
S_charge}) \emph{the} \textquotedblleft minimal coupling
rule\textquotedblright , in that it is \emph{the} \emph{minimum} \emph{%
possible} coupling to EM fields (i.e., a coupling of the charge to the field
that is \emph{linear}, and therefore \emph{the} lowest possible order of
coupling, to the vector potential $\mathbf{A}$, and a coupling that is also 
\emph{linear}, and therefore \emph{the} lowest possible order of coupling,
to the charge $q$) that is permitted by relativity. It can be shown that the
usual \textquotedblleft minimal coupling rule\textquotedblright\ (\ref%
{minimal coupling rule}) then follows from (\ref{Trial solution for S_charge}%
) in the non-relativistic limit in the Hamiltonian formulation of quantum
mechanics.

Now we are in a position to justify the \textquotedblleft Huygens'
construction\textquotedblright\ shown in Figure 1, in which we assume that a
single electron is moving non-relativistically in a magnetic field.

\begin{figure}
\includegraphics[width=5in]{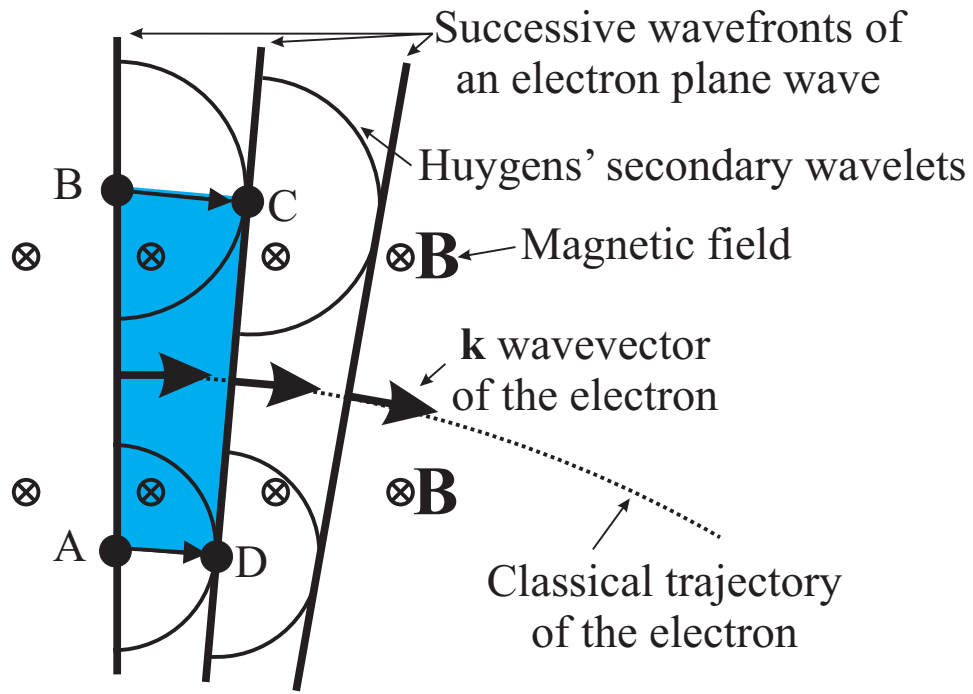}
\caption{Huygens' construction applied
to an electron plane wave propagating towards the right in a magnetic field $%
\mathbf{B}$, which points into the page. The secondary wavelets emitted from
points $A\ $and $B$ on the primary wavefront will suffer a relative
Aharonov-Bohm phase shift arising from the magnetic flux enclosed in the
trapezoid $ABCD$. This phase shift leads to the bending of the normals to
the successive wavefronts, and leads to a curved classical trajectory.}
\end{figure}

When the electron wave
propagates through a space in which there exists a uniform magnetic field $%
\mathbf{B}$ that is applied perpendicularly to the electron's initial
momentum, as in Figure 1, there will arise an Aharonov-Bohm phase shift
between the two secondary Huygens wavelets emanating from points $A$ and $B$
and arriving at points $C$ and $D$, which is given by%
\begin{equation}
\Delta \phi =\frac{1}{\hbar }\oint_{ABCD} \mathbf{A}\cdot d\mathbf{l}%
=\frac{1}{\hbar }e\Phi _{\text{enc}}  \label{A-B phase shift}
\end{equation}%
where $e$ is the electron charge, $\mathbf{A}$ is the vector potential, and $%
\Phi _{\text{enc}}$ is the flux enclosed within the trapezoid $ABCD$ in
Figure 1.

The Aharonov-Bohm phase shift arises here because the electron, as it moves
from $A$ to $D$, accumulates the Dirac phase%
\begin{equation}
\Delta \phi _{AD}=\frac{1}{\hbar }\int_{A}^{D}e\mathbf{A}\cdot d%
\mathbf{l}
\end{equation}%
and as it moves from $B$ to $C$, it accumulates the Dirac phase%
\begin{equation}
\Delta \phi _{BC}=\frac{1}{\hbar }\int_{B}^{C}e\mathbf{A}\cdot d%
\mathbf{l}
\end{equation}%
The difference between these two Dirac phases will then be given by the
closed-path integral (\ref{A-B phase shift})%
\begin{equation}
\Phi _{\text{enc}}=\oint_{ABCD} \mathbf{A}\cdot d\mathbf{l=}%
\iint_{S\left( ABCD\right) } \left( \mathbf{\nabla \times A}\right) \cdot d%
\mathbf{S=}\iint_{S\left( ABCD\right) } \mathbf{B}\cdot d\mathbf{S}
\end{equation}%
which is a manifestly gauge invariant quantity, namely, the enclosed
magnetic flux $\Phi _{\text{enc}}$. It can then be shown that the resulting
\textquotedblleft tilting\textquotedblright\ of the wavefronts or phase
fronts shown in Figure 1 leads to the Lorentz force and a classical
cyclotron orbit.

From the above analysis, it is clear that, in addition to the usual
\textquotedblleft $m\mathbf{v}$\textquotedblright\ momentum (or
\textquotedblleft kinetic momentum\textquotedblright ) that a neutral
particle with mass $m$ would have due its motion through space with a
velocity $\mathbf{v}$, a charged particle with a charge $q$ in the field of
a vector potential $\mathbf{A}$ will acquire an extra piece of
\textquotedblleft electromagnetic momentum\textquotedblright\ due its
charge, which is given by%
\begin{equation}
\mathbf{p}_{\text{charge}}=q\mathbf{A}  \label{charged momentum}
\end{equation}%
even if the charge $q$ is at rest and has no \textquotedblleft $m\mathbf{v}$%
\textquotedblright\ momentum. This \textquotedblleft $q\mathbf{A}$%
-type\textquotedblright\ momentum exists at both the classical level of
description (via the Landau invariance argument) and at the quantum level of
description (via the Feynman--Dirac--Aharonov-Bohm argument).

\begin{figure}
\includegraphics[width=5in]{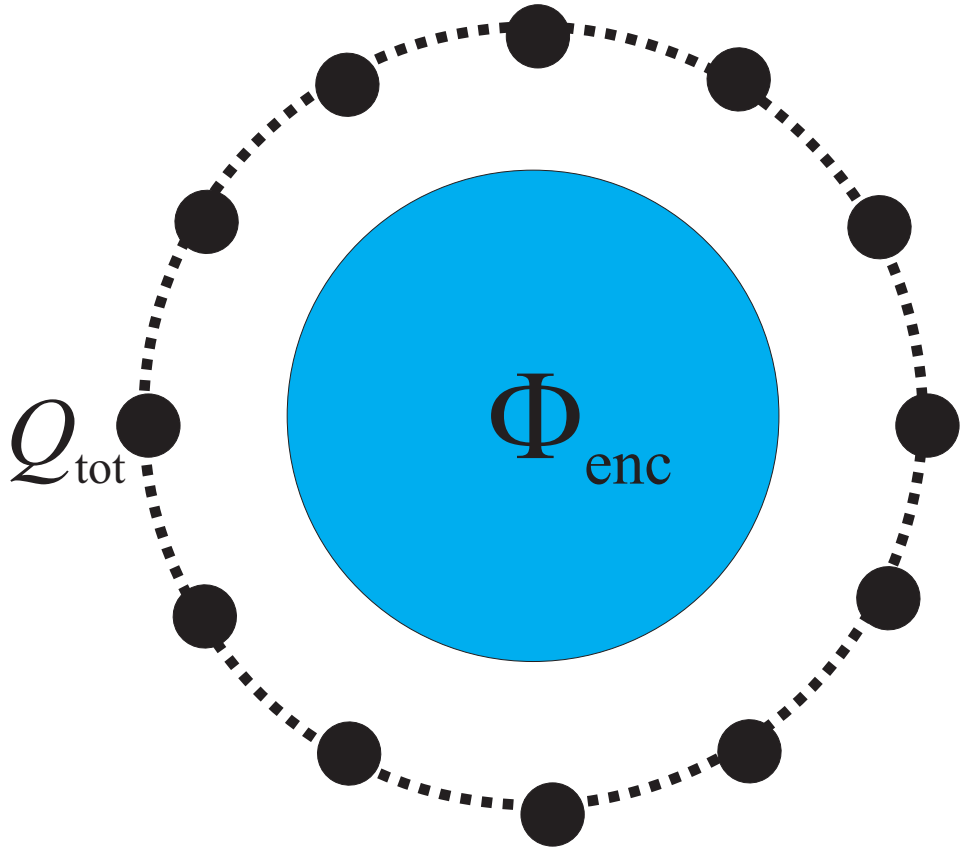}
\caption{A \textquotedblleft
lumpy\textquotedblright\ charged ring of radius $r$ with a total charge of $%
Q_{\text{tot}}$ consisting of a uniform distribution of equally charged
conducting spheres (in \emph{solid black}) with insulation (in \emph{dashed
black}) separating them, forms the rotating member of a torsional pendulum.
This \textquotedblleft lumpy\textquotedblright\ ring of charges $Q_{\text{tot%
}}$ is placed around an infinitely long solenoid containing a flux $\Phi _{%
\text{enc}}$ (in \emph{blue}). The electromagnetic angular momentum of this
system is given by the charge-flux product $Q_{\text{tot}}\Phi _{\text{enc}}$
divided by $2\protect\pi $, even when the charges $Q_{\text{tot}}$ are at
rest. When the charge $Q_{\text{tot}}$ is suddenly discharged and disappears
into the ground, the electromagnetic angular momentum $Q_{\text{tot}}\Phi _{%
\text{enc}}/2\protect\pi $ will suddenly disappear, and the ring must
undergo an angular recoil due to the conservation of angular momentum, and
must start torsionally oscillating around the axis of the solenoid.}
\end{figure}

However, a manifestly gauge-invariant result, even at the purely classical
level, can be gotten by considering a system with \emph{angular} momentum in
a \emph{circularly symmetric} configuration, as illustrated in Figure 2,
where there is a \textquotedblleft lumpy\textquotedblright\ charge
distribution, which is similar to the configuration used in
\textquotedblleft Feynman's paradox\textquotedblright\ \cite{Feynman}. For
then%
\begin{eqnarray}
\oint_{C} \mathbf{p}_{\text{charge}}\cdot d\mathbf{l} &=&\oint_{C} %
q\mathbf{A}\cdot d\mathbf{l=}  \notag \\
\left\langle p_{\text{charge}}\right\rangle \cdot 2\pi r &=&Q_{\text{tot}%
}\Phi _{\text{enc}}
\end{eqnarray}%
\begin{equation}
\therefore L_{\text{charge}}=\frac{1}{2\pi }Q_{\text{tot}}\Phi _{\text{enc}}
\label{A-B angular momentum-a}
\end{equation}%
where $C$ is a circle of radius $r$, where $Q_{\text{tot}}$ is the total
charge, which is the sum of all the discrete charges distributed around the
ring, and where $\Phi _{\text{enc}}$ is the flux enclosed by the
\textquotedblleft lumpy\textquotedblright\ charged ring. Here $\left\langle
p_{\text{charge}}\right\rangle $ denotes the average of the magnitude of $%
\mathbf{p}_{\text{charge}}$ over the ring, and%
\begin{equation}
\oint_{C} q\mathbf{A}\cdot d\mathbf{l}=Q_{\text{tot}}\left\langle \oint_{C} %
\mathbf{A}\cdot d\mathbf{l}\right\rangle \mathbf{=}Q_{\text{tot}}\Phi _{%
\text{enc}}
\end{equation}%
denotes an average over the \textquotedblleft lumpy\textquotedblright\
charge distribution of the quantity $q\mathbf{A}$, as if there were a
smooth, continuous, uniform distribution of bound charges frozen inside an
insulating ring, in a continuum model of the \textquotedblleft lumpy\textquotedblright\ charge ring.

The \emph{classical} expression%
\begin{equation}
L_{\text{charge}}=\frac{1}{2\pi }Q_{\text{tot}}\Phi _{\text{enc}}
\label{classical expression}
\end{equation}%
obtained in (\ref{A-B angular momentum-a}) is in fact the same as that for
the Aharonov-Bohm angular momentum at the \emph{quantum} level, when one
substitutes $Q_{\text{tot}}=e$ for the charge of a single, completely
delocalized electron, which possesses a perfect quantum phase coherence
around the ring. Note, however, that there is no requirement in the
classical case for the discrete, \textquotedblleft lumpy\textquotedblright\
charge distribution illustrated in Figure 2 to possess a single, coherent
quantum-mechanical phase over the entire ring, such as is required in the
case of the Aharonov-Bohm effect. Each macroscopic \textquotedblleft
lump\textquotedblright\ of charge may have decohered, and may have become a
completely localized lump of classical matter which is at a perfect rest.
Therefore one concludes that this \textquotedblleft $q\mathbf{A}$%
-type\textquotedblright\ of \emph{angular} momentum should also exist at the 
\emph{classical}, macroscopic level of description, as well as at the \emph{%
quantum} level. We shall call this classical type of angular momentum
\textquotedblleft Aharonov-Bohm-like\textquotedblright\ angular momentum.

An additional classical argument for the existence of the classical,
\textquotedblleft Aharonov-Bohm-like\textquotedblright\ angular momentum can
be gotten by applying Faraday's law of induction to Figure 2. Suppose that
one were to increase the flux from zero inside the solenoid linearly with
time by ramping up the current flowing through its coils. Then by Faraday's
law%
\begin{equation}
\oint_{C} \mathbf{E}\cdot d\mathbf{l=-}\frac{d\Phi _{\text{enc}}}{dt}
\end{equation}%
Therefore the magnitude of the resulting azimuthal electric field $E$
evaluated on the circle of radius $r$ will be given by%
\begin{equation}
E\cdot 2\pi r\mathbf{=}\frac{d\Phi _{\text{enc}}}{dt}
\end{equation}%
This will lead to a torque $\tau $ on the distribution of charges on the
\textquotedblleft lumpy\textquotedblright\ ring around its axis with a
magnitude%
\begin{equation}
\tau =Fr=(Q_{\text{tot}}E)r\mathbf{=}\frac{Q_{\text{tot}}}{2\pi }\frac{d\Phi
_{\text{enc}}}{dt}
\end{equation}%
Integrating the torque over time to obtain the electromagnetic angular
momentum $L_{\text{charge}}$ of the \textquotedblleft
lumpy\textquotedblright\ charged ring, one finds that%
\begin{equation}
L_{\text{charge}}=\int \tau dt=\frac{1}{2\pi }Q_{\text{tot}}\Phi _{\text{enc%
}}
\end{equation}%
in agreement with (\ref{classical expression}). Note that this Faraday-law argument holds at the \emph{classical} level.

This suggests that there may exist manifestations of \textquotedblleft
Aharonov-Bohm-like\textquotedblright\ \emph{nonlocality} in \emph{classical}
experiments, whenever the charge $Q_{\text{tot}}$ exists in a disjoint
region of space from the flux $\Phi _{\text{enc}}$. Ultimately, we should
test this idea out experimentally in a Tonomura-type experiment, where the
space which contains the \textquotedblleft lumpy\textquotedblright\ circular
ring of distributed charges that sum to a total charge $Q_{\text{tot}}$, and
the space which contains the enclosed magnetic flux $\Phi _{enc}$,\ are
separated into two \emph{mutually exclusive} regions, such as the two
separated regions of space within two linked tori.

However, we should approach this final Tonomura-type experiment in stages.
In a first experiment, there would be no attempt to separate between the
regions of the electric fields of $Q_{\text{tot}}$ and of the magnetic
fields of $\Phi _{\text{enc}}$, so that the recoil can be explained entirely
based on the \emph{classical} torque arising from the Lorentz force acting
on the discharging currents along the spokes of a charged wheel patterned
after Figure 2 \cite{Spokes-of-wheel}. The solenoid of Figure 2 would be
replaced by a permanent magnet with exposed pole faces in this first
experiment. Then in the second experiment we could attempt to shield the
discharge currents by using adjacent ground planes in a microstrip
configuration, so that the Lorentz force on the discharging currents in the
spokes of the wheel is cancelled out by the Lorentz force on the
image-charge counter-currents in the ground plane. Finally, in the third
experiment, we could use both ground planes and mu-metal shields in
conjunction with a toroidal configuration of the magnetic flux trapped
inside a high permeability metallic path, in an attempt to separate the
electric and magnetic fields more or less completely into two \emph{mutually
exclusive} regions of space. All of these macroscopic, classical experiments could be done at room temperature.

\pagebreak

\begin{equation}
F=qvB
\end{equation}%
acting upon each spoke. Dividing a given spoke into infinitesimal segments
of length $dr$, the torque $\tau $ per spoke exerted on the central axle of
the wheel will be given by%
\begin{equation}
\tau =\frac{1}{a}\int_{0}^{a}Frdr=\frac{1}{2}Fa=\frac{1}{2}qvBa
\end{equation}%
The angular momentum $L$ imparted to the wheel per spoke by the torque $\tau 
$ during the discharge will therefore be given by%
\begin{eqnarray}
L &=&\int_{0}^{\infty }\tau dt=\frac{1}{2}Baq\int_{0}^{%
\infty }vdt  \notag \\
&=&\frac{1}{2}Baq\int_{0}^{\infty }\frac{dx}{dt}dt=\frac{1}{2}%
Baq\int_{0}^{a}dx  \notag \\
&=&\frac{1}{2}Ba^{2}q=\frac{1}{2\pi }B\pi a^{2}q=\frac{1}{2\pi }\Phi _{\text{%
enc}}q
\end{eqnarray}
where $v=dx/dt$ is the velocity of the pulse of charge $q$ moving along a
given spoke, and where $\Phi _{\text{enc}}=B\pi a^{2}$ is the magnetic flux
enclosed by the wheel of radius $a$. Summing over all the charges on the
wheel, one obtains that the total final (recoil) angular momentum of the
wheel will be given by 
\begin{equation}
L_{\text{total charge}}=\frac{1}{2\pi }\Phi _{\text{enc}}Q_{\text{tot}}
\end{equation}%
where $Q_{\text{tot}}$ is the total charge on the wheel. Note again that
this derivation has been entirely performed at the classical level. This is
in agreement with the more general classical result (\ref{classical
expression}) obtained via (\ref{A-B angular momentum-a}).


\begin{thebibliography}{9}
\bibitem{Spokes-of-wheel} Imagine that the charged, conducting spheres in
Figure 2 were connected via radial wires to a charged central conductor,
which acts as the axle of a \textquotedblleft lumpy\textquotedblright\
charged wheel of radius $a$ of a torsional pendulum. These radial wires can
then function as the spokes of the wheel. Moreover, imagine that the
solenoid of Figure 2 were to be replaced by a permanent magnet whose radius
is also $a$, and whose pole face is concentrically placed in close proximity
underneath the wheel, but not touching it, so that the wheel can turn
freely. As a result, there will be a small gap through which the vertical,
uniform magnetic induction $B$ of the magnet will be applied to each spoke
of the wheel. Now let each sphere initially contain a charge $q$. When the
central conductor is suddenly connected to ground at $t=0$, a discharging
pulse of charge $q$ will travel radially with some velocity $v~$along each
spoke of the wheel in the presence of $B$ towards the grounded central axis,
so that there will be an azimuthal component of the Lorentz force of strength%
\end{thebibliography}
\end{document}